\documentclass[amssymb,amsmath,nobibnotes,aps,prl,showpacs, twocolumn]{revtex4}
\usepackage{docs}%
\usepackage{bm}%
\usepackage{graphicx}
\expandafter\ifx\csname package@font\endcsname\relax\else
 \expandafter\expandafter
 \expandafter\usepackage
 \expandafter\expandafter
 \expandafter{\csname package@font\endcsname}%
\fi
\def\aa{\AA\,} 
\def\Journal#1#2#3#4{\textit{#1} \textbf {#2}, \textsl{#3}, #4} 
 
\def\JAP{ J. Appl. Phys.}

\def\PRL{ Phys. Rev. Lett.}

\def\PRB{{ Phys. Rev.} B}

\def\SAB{ Sens. Act. B: Chemical} 
\begin{document}

\title{Charge transport in purple membrane monolayers: A sequential tunneling approach}

\author{E. Alfinito\footnote{Corresponding author}}%
\email{eleonora.alfinito@unisalento.it}
\author{J.-F. Millithaler}%
\email{jf.millithaler@unile.it}
\author{L. Reggiani}
\email{lino.reggiani@unisalento.it}
\affiliation{Dipartimento di Ingegneria dell'Innovazione. Universit\`a del Salento,
via Monteroni, I-73100 Lecce, Italy, EU\\
CNISM - via della Vasca Navale, 84, I-00146, Roma, Italy, EU }

\date{\today}%
\begin{abstract}
Current voltage (I-V) characteristics in proteins can be sensitive to conformational change induced by an external stimulus (photon, odour, etc.). This sensitivity can be used in medical and industrial applications besides shedding  new light in the microscopic structure of  biological materials.
Here, we show that a sequential tunneling model of carrier transfer between neighbouring amino-acids in a single protein can be the basic mechanism responsible of  the electrical properties measured in a wide range of applied potentials. 
We also show that such a strict correlation between the protein structure and the electrical response can lead to a new generation of nanobiosensors that mimic the sensorial activity of living species.   
To demonstrate the potential usefulness of protein electrical properties, we provide a microscopic interpretation of  recent I-V experiments carried out in bacteriorhodopsin at a nanoscale length.
\end{abstract}
%\keywords{receptor proteins, electron transport, Fowler-Nordheim equations}
\maketitle
\vspace{-0.5cm}
\qquad \qquad{\small Keywords: bacteriorhodopsin, electron transport,
\noindent Fowler-Nordheim
tunneling}
%\section{Table of contents}

The nature of electron transport (ET) in biological material is matter of outstanding interest both for the pure speculative aspects and for the applications.
As a matter of fact, organic/biological-based devices are the new frontier of technology, due to their potential low cost, low size, high specificity, etc. \cite{Bond, Hou}. 
Usually biological  materials are not easy to be investigated, because a standard way of preparation is still not available \cite{Hou}.
Nevertheless, there are some relevant exceptions, like monolayers of purple membrane (PM), a part of the cell membrane of the halophile \textit{Halobacterium salinarum}, which is easy to prepare and suitable for direct measurements. 
PM is constituted by a single type of protein, the light receptor bacteriorhodopsin (bR), organized in trimers and stabilized by lipids \cite{Corcelli}.
The entire structure appears as a 2D hexagonal crystal lattice.
The natural role of bR is to use sun light for pumping protons outside the cell and in doing so, it changes its tertiary structure (conformational change). 
\par
Recently,  current-voltage (I-V)  characteristics of purple membrane were analyzed under different experimental conditions \cite{Jin,Gomila,Ron}, giving clear evidence  of super-Ohmic and illumination-dependent responses \cite{Jin}.
These measurements are of paramount importance for the understanding of the mechanism of ET, also in relation with the protein
activation, furthermore, they are very promising
for the development of a new generation of organic based devices \cite{Kannan}. 
The seminal  experiment \cite{Jin} was  carried on metal-insulator-metal (MIM) junctions of millimetric diameter,
with the insulator constituted by a 5 nm  monolayer of PM. 
The measurements covered a small range of bias ($0 \div 1$ V), because of
the high value of current response (nA level).
The response is found to be slightly super-Ohmic and grows overall a factor of 2 when the sample is irradiated by a green light.
These results suggest that in this protein, like in some
organic polymers \cite{Zvy},  ET is ruled by tunneling mechanisms. 
Furthermore, the large thickness of PM (5 nm) strongly proposes
%the strong dependence of the current from the presence/absence of the cromophore 
the possibility of multiple carrier jumps across the protein (\textit{sequential} tunneling) \cite{Jin, Ron}.
\par
In a later experiment\cite{Gomila}, the I-V characterization was performed at the \textit{nanometric} scale, with the technique of c-AFM (conductive atomic force microscopy).
Accordingly, one of the contacts is constituted by the tip of the c-AFM
($100 \div 200$ nm of nominal radius). 
With respect to the first measurement, in the
common bias range, the current response is lower for about 4 order of magnitude, thus the sample is able to sustain higher
voltage up to about $5 \div 10$ V.
Measurements were performed without any extra light irradiation and with different tip indentations from about 4.6 nm down to 1.2 nm.
At voltages above about 2 V, the presence of a cross-over between
the direct tunneling regime and the injection or Fowler-Nordheim tunneling regime was evidenced.
\par
At present, a microscopic interpretation of the above experiments is still in its infancy and,  apart from some attempts \cite{Gomila,Epl}, a unifying approach able to  explain the main features and possibly to be predictive for analogous physical systems is lacking. 
This paper aims to fill this lack of knowledge by implementing  a new ET model, hereafter called INPA (impedance network protein analogue). INPA uses a microscopic  description of the protein tertiary structure that takes the amino acids as single centers of interaction, responsible of charge transfer.
Finally, the  charge transport through a single protein results as a process due to the simultaneous activation of multiple pathways in an \textit{impedance network} on the line of the tunnelling pathway method developed by Onuchic and cowkers \cite{Beratan}. 
The network approach has also strong analogies with a percolative process \cite{pennetta00} and thus the solution of the transport problem is carried out within a stochastic method.
With respect to existing approaches\cite{Adam, Nitzran}, INPA  has some advantages:
$i)$ the tertiary structure of the protein is directly correlated to the macroscopic observables;
$ii)$  the different electrical responses in the presence/absence of a green light is reproduced;
$iii)$  the interpretation of different data \cite{Jin,Gomila} is reconciled   
$iv)$ it can be applied to other proteins whose tertiary structure is known. 
\par
In brief the layout of the INPA is as follows.
The protein structure is coarse-grained described by means of the C$_\alpha$ positions, as obtained by the protein
data base (PDB) or homology modeling\cite{Epl}. 
Each C$_\alpha$ position is taken as corresponding to a node in a graph whose links describe the electrical interactions between  amino-acids.
The degree of proximity for each node is assigned by a cut-off interaction radius, say $R_{C}$. 
In the present analysis, we choose $R_{C}$ = 6 \aa, a value that optimizes the native to activated state resolution \cite{Epl}.
Each link is associated with an impedance (a simple resistance in this case)  whose value depends on the distance between amino-acids. In particular:
$r_{i,j}=\rho\, l_{i,j}/\mathcal{A}_{i,j}$, 
where $\rho$ indicates the resistivity here taken to be same for all the links; $\rho$, in general, depends on the voltage as detailed below, the pedices $i,j$ refer to the amino-acids between which the link is stretched,
$l_{i,j}$ is the distance between the labeled aminoacids taken as point like centers  and ${\mathcal{A}_{i,j}}$ is the cross-sectional area shared by the labelled amino-acids:
 ${\mathcal{A}_{i,j}}= \pi\left(\textsl{ R}_C^{2}-l^{2}_{i,j}/4\right)$.
The graph is thus mapped into an impedance network.
\noindent 
\begin{figure}[htb]
\centering\noindent
\includegraphics[width=15pc]{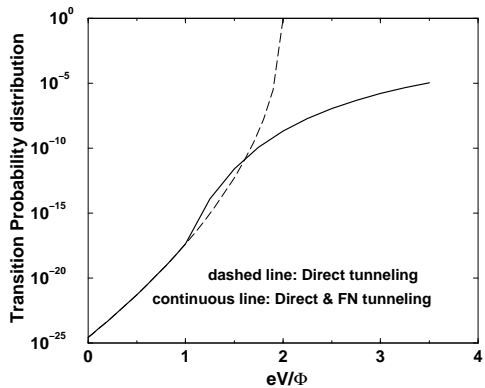}
\caption{Transmission probability for the reported two tunneling regimes}
\label{fig:1}
\end{figure}
To take into account the superlinear response, $\rho$ is chosen to depend on the voltage drop  as:
\begin{equation}
\rho(V)=\left\{\begin{array}{lll}
\rho_{MAX}& \hspace{.5cm }& eV \leq  \Phi  \\ \\
 \rho_{MAX} (\frac{\Phi}{eV})+\rho_{min}(1- \frac{\Phi}{eV}) &\hspace{.5cm} & eV \ge  \Phi 
 \end{array}
  \right.
\label{eq:3}
\end{equation}
where $\rho_{MAX}$ is the resistivity value which should be used to fit the I-V characteristic at the lowest voltages,  $\rho_{min} \ll \rho_{MAX}$  plays the role of an extremely low series resistance, limiting the current at the highest voltages,  and 
$\Phi$  is the threshold energy separating the two tunneling regimes (a kind of effective height of a tunneling barrier).
Since ET is here interpreted in terms of a sequential tunneling between neighbouring amino-acids, the above interpolation formula reflects the different voltage dependence in the prefactor of the current expression \cite{Wang}: $I\sim V$ in the direct tunneling regime, and $I\sim V^2$ in the FN tunneling regime.
\par
For  the transmission probability of the tunneling mechanism we take the expression given by Ref. \cite{Gomila, Simmons}:
\begin{equation} 
\mathcal{P}^{D}_{i,j}= \exp \left[- \frac{2 l_{i,j}}{\hbar} \sqrt{2m(\Phi-\frac{1}{2}
eV_{i,j})} \right] \ ,
\hspace{0.7cm}
 eV_{i,j}  \leq \Phi  \,
\label{eq:1}
\end{equation}
\begin{equation}\label{eq:2}
\mathcal{P}^{FN}_{ij}=\exp \left[-\left(\frac{2l_{i,j}\sqrt{2m}}{\hbar}\right)\frac{\Phi}{eV_{i,j}}\sqrt{\frac{\Phi}{2}} \right] \ , 
\hspace{0.7cm}
 eV_{i,j} \ge \Phi  \ ;
\end{equation}
where $V_{i,j}$ is the local potential drop between the couple of $i,j$ amino-acids and $m$  is the electron effective mass, here taken the same of the bare value. 
\par
Figure 1 reports the shape of the tunneling transmission probability that  includes  both the direct and injection regimes (continuous curve) together with that corresponding to direct tunneling only (dashed curve). 

\par
In the solution of the  resistor network, the tunneling mechanism is accounted for by the following procedure.
First, the network is electrically solved by using the value $\rho_{MAX}$ for all the elemental resistances.
Second, by using a Monte Carlo acceptance-rejection procedure, each $\rho_{MAX}$ is stochastically replaced  by $\rho_{min}$  using the probability in Eq. (\ref{eq:1}) and Eq. (\ref{eq:2}) according to the local potential drops calculated in the first step. 
In the high voltage region ($eV_{i,j} > \Phi $), if the stochastic procedure gives a rejection, then $\rho_{MAX}$ is replaced by $\rho(V_{i,j})$ of  Eq. (\ref{eq:3}). 
The network is then electrically updated with the new distributed values of $\rho(V_{i,j})$.
Third, the electrical update is iterated (typically $10^{6} - 10^{8}$  iterations depending on the value of the applied voltage) by repeating the second step until the value of the  network associated  current, converges within an uncertainty less than a few per cent.
\noindent 
\begin{figure}[htb]
\centering\noindent
\includegraphics[width=16pc]{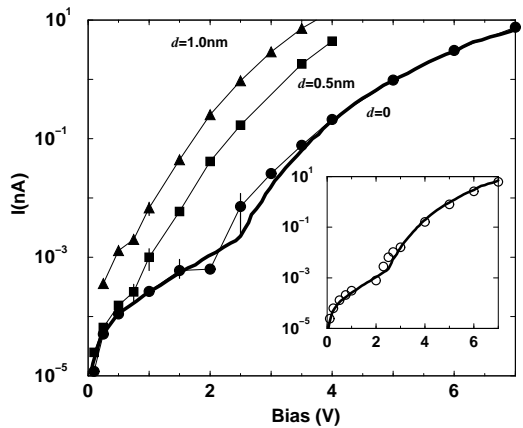}
\caption{I-V characteristics obtained by simulations with extended contacts at the different indentations.
The symbols and the thin dashed curves refer to calculations. 
The thick continuous line refer to the experimental data in the absence of indentation (d=0) when the electrode distance is 4.6 nm \cite{Gomila}. 
In the inset the theoretical fit for d=0 reported by symbols is performed with point-like contacts (see text).}

\label{fig:2}
\end{figure}
\begin{figure}[htb]
%\begin{minipage}[t]{12pc}
\includegraphics[width=16pc]{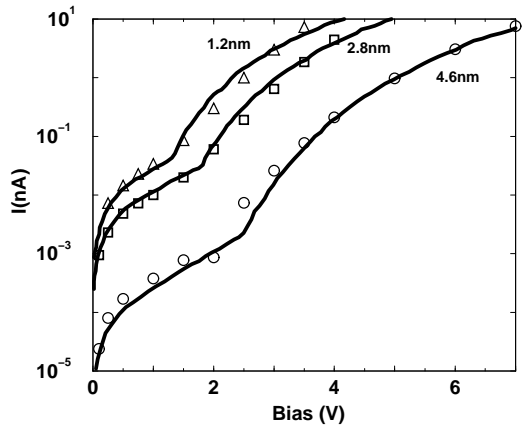}
%\end{minipage}\hspace{7.5pc}%
%\begin{minipage}[t]{14pc}
%\includegraphics[width=14pc]{logfittot.eps}
%\end{minipage}\hspace{1.5pc}%
\caption{I-V characteristics obtained by simulations with extended contacts at the different indentations including a leakage current. 
The thick continuous lines refer to the experimental data \cite{Gomila}
and the symbols to numerical simulations.}
\label{fig:3}
\end{figure}
\begin{figure}[htb]
\centering\noindent
\includegraphics[width=16pc]{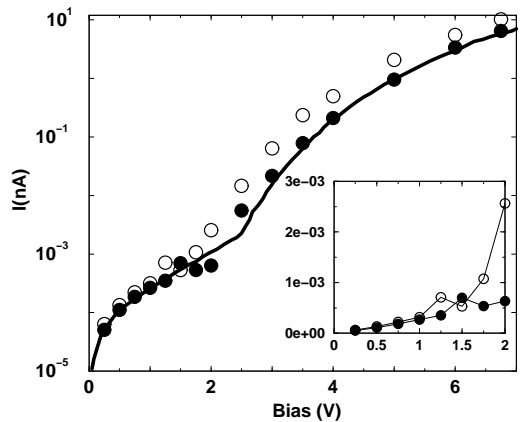}
\caption{I-V characteristics for the native and activated states of bR
performed with point-like contacts. 
Closed and open symbols refer to native and activated states, respectively. Thick curve to experiments for an electrode distance L=4.6 nm \cite{Gomila}.
In the inset the current is rescaled to reproduce the value of the native state at 1 V measured in Ref. (\cite{Jin}).}
\label{fig:4}
\end{figure}
\par
To model bacteriorhodopsin in its native  state (in dark) 
\cite{Epl} we have taken the PDB entry 2NTU. 
The network is then studied with two different contact configurations: 
point-like contacts (in line with previous works), and extended contacts. 
In the first configuration, the network is connected to the external bias by means of perfectly conductive contacts, the input on the first amino-acid, the output on the last amino-acid of the primary structure. 
In the second configuration, the input simulates the extended tip of the AFM device. 
Accordingly, all the nodes with the z-coordinate (direction of the tip penetration) larger than that of the first amino acid, assume the same potential value.
The output, point-like, remains on the last amino acid. 

\par
The values of $\rho_{MAX}$, $\rho_{min}$, and $\Phi$  are obtained by fitting the experiments of \cite{Gomila}
corresponding to an electrode distance of $L=4.6$ nm.
For this distance, the measurements are associated with a current crossing a single layer of proteins.  
Actually, we found: $\rho_{MAX} = 4\times 10^{13}\,\Omega$ \aa\, $\rho_{min} =  4 \times 10^{5}\,\Omega$ \aa  and $\Phi = 219$  meV. 
The large difference between  the $\rho_{MAX}$ and the $ \rho_{min}$ values is dictated by the six
order of magnitude spanned by the current values.
The values of $\rho_{MAX}$, $\rho_{min}$  correspond to the
the macroscopic  resistance given by the  $V/I$ ratio of the
experiment. 
In such a view,  the current measured at 1 V in Ref. \cite{Jin} yields a number of trimers equal to about $N= 10^9$ for a sample area of $2 \times 10^{11} \ nm^2$, thus to a resistivity of about $10^{20}\ \Omega$ \aa, for trimer \cite{Epl}. 
By assuming, in present case, the same trimer resistivity, we  estimate
a number of trimers involved in the measured current of about  $N= 10^6$, thus leading to a crossing  area of  about $ 10^{7} nm^{2}$.
However, within a MIM electrical analogue, in \cite{Gomila} the effective area deduced by the fit was found  to be of   $0.1  \times \ nm^{2}$, about eight order of magnitude smaller than what estimated above.
This dramatic difference is here mainly attributed to the MIM electrical analogue used in \cite{Gomila} that contrasts with the sequential tunneling model used here.
\par
Figure ~\ref{fig:2} compares the numerical  and experimental data for
the extended contact model and the point-like contact model at L=4.6 nm (in the inset).
In both the cases, the agreement is  within the experimental and numerical uncertainty, and thus considered to be satisfactory.
The threshold energy $\Phi = 219 \ meV$, is taken to be  independent of the contact choice. 
On the other hand, when  going from the point-like to the extended contact configuration the the fitting values of $\rho_{MAX}$ increases from $4\times 10^{13}\,\Omega$ \AA\, to $8\times 10^{13}\,\Omega$ \AA\, and also those of $\rho_{min}$ increases from $4 \times 10^{5}\,\Omega$ \AA\, to $4 \times 10^{6}\,\Omega$ \AA. 
\par
The position of the  extended contact is then changed in order to reproduce the experiments  obtained at  different indentation of the tip.
At increasing depths of the extended contact, the net effect is a reduced number of  amino acids involved in the electrical transport. 
As a consequence, higher currents and a shift to lower potential values for the crossover between the direct and the FN regime is expected.
Numerical calculations reported in Fig. {\ref{fig:2}} confirm these expectations.
In particular, the flat contact that simulates the tip indentation was placed at the depths of 0.50 nm, and 1.0 nm, from the top of the protein.
For these depths, the experimental data at increasing tip indentations exhibit a quantitative agreement  with the modelling in the region of  high voltages where FN tunneling prevails.
Even if the experimental indentation of the tip is greater for about a factor of 3 with respect to that of the simulations, we consider the agreement between theory and experiments to be satisfactory in view of the simplifications needed to convert the single protein calculation into
the macroscopic value measured.
However, at low voltages the results of the simulations underestimate for up to an order of magnitude the values of experiments.
The  disagreement at low voltages is here overcome by assuming the existence of a leakage contribution,  
probably associated with the complexity of the contact regions, constituted by trimers and lipids, with respect to the used model \cite{Bartlett}. 
Accordingly,  the single resistance associated with each link is replaced by the parallel of two resistances, 
one pertaining to the protein and the other to a more realistic modelling of the contact regions.  
\par
Figure (\ref{fig:3}) reports the currents for different indentations when
an Ohmic leakage current is added to the values reported  in Fig. 2.  
The best fit is obtained by taking for the leakage
resistance the values: $8.4 \times 10^{12} \  \Omega$, $0.11 \times 10^{12} \  \Omega$,  $0.036 \times 10^{12} \ \Omega$, respectively for L values  of  $4.6$, $2.8$ and $1.2$ nm. 
The leakage resistance at L=4.6 nm is taken equal to the protein resistance value at low bias.
The decrease of the leakage resistance at increasing indentation can be related to geometrical effects associated with  the decreasing of the inter-electrode distance  and the increasing surface of the tip contact when penetrating the protein.  
Interestingly enough, the microscopic interpretation is obtained with a length-independent electron effective mass, contrary to the case of  Ref. \cite{Gomila}  where to fit experiments the carrier effective mass should increase over one order of magnitude at increasing the indentation. 
(We notice that a larger (smaller) value of the effective mass would produce a shift to lower (higher) value of the transmission probability
which can be compensated by a corresponding change in the value of the energy threshold without significant modifications of the microscopic interpretation.)
Most important, the value of the energy threshold is reduced, for about a factor of ten,  when compared with the values of the barrier height found in  Ref. \cite{Gomila}. 
Both these features are a consequence of the sequential tunneling mechanism assumed here, and that replaces the single tunneling mechanisms of the MIM model previously used \cite{Gomila}.

\par
As above anticipated, the INPA is able to interpret the behaviour of the I-V characteristics carried out with an 
\textit{electrode-bilayer-electrode} structure and when the protein is illuminated or less by green light  \cite{Jin}.
To this purpose, the model is applied to the PDB entry 2NTW (describing the activated state of bR) \cite{Epl} and the
data, calculated with the same  parameters used for calculating the I-V characteristic of the native state, in the extended contact configuration. 
Figure ~\ref{fig:4} reports the simulated data calculated for both the native and activated state. 
The trend evidenced by experiments is here reproduced without introducing arbitrary parameters.  
We remark, that present results are compatible with those reported by some of the Authors in Ref. \cite{Epl}. 
Here, the higher value of $\Phi$ used to fit the data
of Ref. \cite{Gomila} leads to some minor differences in the current responses of the native and activated states which should be justified by the complexity of physical system investigated.
\par
In conclusion, we propose a sequential tunneling mechanism for charge transport in bacteriorhodopsin.
The model permits a consistent interpretation of a set of experiments carried out in a wide range of applied electrical potentials and in the presence or less of an external green light.
The tertiary structure of the protein enters as direct data input and enables one to relate quantitatively conformational change and sensing action of the protein.
Finally, also data obtained in very stressing conditions, like the penetration of an AFM  tip in the protein membrane, can be finely reproduced.
The qualitative and quantitative agreement between the numerical results and experiments  poses the INPA,  implemented here for a sequential tunneling mechanism,  as a physical plausible model to  investigate the electrical properties in other  proteins pertaining to the transmembrane family.

This research is supported by the European Commission under the Bioelectronic Olfactory Neuron Device (BOND) project within the grant agreement number 228685-2.

\end{document}